\documentstyle[12pt,A4]{article}
\makeatletter
\def\warnred#1#2{\@ifundefined{#1}{}{\@warning{REDEFINING \expandafter\string
\csname #1\endcsname\space TO \string#2}}}
\warnred{nc}{newcommand}\let\nc\newcommand
\nc\mycomment[1]{\marginpar{\vspace*{-1.5\baselineskip}
\tiny\begin{flushleft}#1\end{flushleft}}}
\nc\REMARK[1]{}
\nc\com{~,\hs}
\nc\hs{\hspace*{2em}}
\nc\mypar{\par\vskip 1ex}
\nc\noi{\noindent}
\nc\xvspace{\par~\par}
\nc\hr{\xvspace\hrule\xvspace}
\nc\ms[1]{\rule{0pt}{#1}}
\nc\SR[2]{\rule[#1]{0pt}{#2}}
\nc\vs[1]{\SR{-#1 mm}{#1 mm}}
\nc\vt[1]{\rule{0mm}{#1mm}}
\nc\vu[1]{\SR{#1 mm}{#1 mm}}
\nc\vv[1]{\SR{#1 mm}{1 mm}}
\nc\initeqn{
\def\theequation{\thesection.\arabic{equation}}\@addtoreset{equation}{section}}
\nc\sect{\mynewpage\initeqn\section}
\nc\mynewpage{}
\warnred{ul}{underbar}
\nc\bea{\begin{eqnarray}}
\nc\be{\begin{equation}}
\nc\bex{\begin{equation}\let\\=\com\catcode`\&=10\relax}
\nc\ee{\end{equation}}
\nc\ena{\end{eqnarray}}
\nc\bit{\begin{itemize}}
\nc\eit{\end{itemize}}
\nc\ba{\begin{array}}
\nc\ea{\end{array}}
\nc\frenchdefs[9]{
\hyphenation{ana-ly-tique ana-ly-se ana-ly-ser ca-no-nique ca-no-niques
coup-lage coup-lages de-scrip-tion diffe-rent diffe-rents diffe-rente
diffe-rentes en-cou-rage-ment en-cou-rage en-cou-rage-ments exemple
exo-tique exo-tiques fer-mio-ni-que fer-mio-ni-ques in-va-riant
in-va-riante jeune jeunes mo-di-fie mo-di-fies mo-di-fiees
in-va-riantes Karls-ruhe Le-gen-dre par-ti-cule par-ti-cules
re-nor-mali-sable re-nor-mali-sables super-espace super-champs
syme-trie syme-tries super-gravite va-ria-tion-nel va-ria-tion-nels
va-ria-tion-nelle va-ria-tion-nelles veri-fier veri-fie veri-fiant
veri-fient}
\nc\apparai[1]{appara{\^\i}#1}
\nc\cad{c'est-\`a-dire}
\nc\coder[1]{d\'e\-ri\-v\'ee#1 co\-va\-ri\-an\-te#1}
\nc\cotrans[1]{\susytrans{#1} covariante#1}
\nc\diff{diff\'e\-ren\-tiel}
\nc\diffop[1]{op\'erateur#1 \diff{#1}}
\nc\diffgeo{g\'eo\-m\'e\-trie \diff le}
\nc\eqref[1]{{\mbox{\'eq. (\protect\ref{#1})}}}
\nc\eqn{\'equation}
\nc\bewgl[1]{\eqn{#1} de mouvement}
\nc\idb[1]{identit\'e#1 de Bianchi}
\nc\MS{Mo\-d\`ele Standard}
\nc\MSS{\MS\ Su\-per\-sy\-m\'e\-tri\-que}
\nc\puisque{\hs\mbox{puisque }}
\nc\scc{super\-champ chiral}
\nc\sccs{super\-champs chiraux}
\nc\susy{super\-sym\'e\-tri}
\nc\susyalg{alg\`ebre de \susy e}
\nc\susytrans[1]{trans\-for\-ma\-tion#1 de \susy e}
\nc\vev[1]{va\-leur#1 moyenne#1 dans le vide}
\nc\wfaktor{{\kappa}}
\nc\mysect[3]{\REMARK{\newpage~\vfill}\sect{#1}\REMARK{\vfill{\Huge$$ #2 $$}\vfill}#3\REMARK{\newpage}}
\warnred{@oldhat}{\string\^ --- This is bad !!!}\let\@oldhat=\^
\@warning{REDEFINING \string\^ FOR \string\^i to work (@oldhat=old def)}
\def\^{\@ifnextchar i{\@oldhat\i\@gobble}{\@oldhat}}
}
\nc\savebot[1]{\addtolength\textheight{#1}}
\nc\savetop[1]{\savebot{#1}\advance\voffset-#1}
\nc\wider[1]{\advance\hoffset -#1 \advance\textwidth #1 \advance\textwidth #1}
\nc\mydraft{\def\mynewpage{\vfill\pagebreak}
	\input fullpage.sty \myheadings \savebot{2cm}}
\nc\mytwocol{\advance\columnsep 1em\twocolumn\parindent0pt}
\@ifundefined{greektex}{}{\endinput}
\newcommand\greeknewcommand[2]{\def#1{{#2}}}
\greeknewcommand{\al}{{\alpha}}
\greeknewcommand{\gm}{{\gamma}}
\greeknewcommand{\dt}{{\delta}}
\def\eps{{\epsilon}}
\greeknewcommand\veps{\varepsilon}
\greeknewcommand{\la}{{\lambda}}
\def\si{{\sigma}}
\greeknewcommand\vk{\varkappa}
\def\vp{\varphi}
\greeknewcommand{\th}{\theta}
\greeknewcommand{\om}{{\omega}}
\greeknewcommand{\ze}{\zeta}

\greeknewcommand{\Gm}{\Gamma}
\greeknewcommand{\Si}{\Sigma}
\greeknewcommand{\Dt}{\Delta}
\greeknewcommand{\La}{\Lambda}
\greeknewcommand{\Th}{\Theta}
\greeknewcommand{\Om}{{\Omega}}

\def\bt{{\bar\theta}}

\greeknewcommand{\bze}{\bar\zeta}
\greeknewcommand{\thb}{\bar\theta}
\greeknewcommand{\vpb}{\bar\vp}
\greeknewcommand\bvp{\bar\varphi}%
\greeknewcommand{\sib}{\bar\sigma}
\greeknewcommand{\lab}{\bar\lambda}
\greeknewcommand{\xib}{\bar\xi}
\greeknewcommand{\psib}{\bar\psi}
\greeknewcommand{\chib}{\bar\chi}
\greeknewcommand{\phib}{\bar\phi}
\greeknewcommand{\zeb}{\bar\zeta}

\greeknewcommand{\Phib}{\bar\Phi}
\greeknewcommand{\Psib}{\bar\Psi}
\greeknewcommand{\Thb}{\bar\Theta}
\greeknewcommand{\Pib}{\overline\Pi}
\greeknewcommand{\Lab}{\bar\Lambda}
\greeknewcommand\Sib{\bar\Si}
\greeknewcommand\Xib{\bar\Xi}

\greeknewcommand{\da}{{\dot{\alpha}}}
\greeknewcommand{\db}{{\dot{\beta}}}
\greeknewcommand{\dd}{{\dot{\delta}}}
\greeknewcommand{\dep}{{\dot{\epsilon}}}
\greeknewcommand{\de}{{\dot\varepsilon}}
\greeknewcommand{\dg}{{\dot{\gamma}}}
\greeknewcommand{\dm}{{\dot{\mu}}}

\greeknewcommand{\dv}{\dot\varphi}

\greeknewcommand{\hcq}{{\hat{\cal Q}}}
\greeknewcommand{\hB}{\hat{B}}
\greeknewcommand{\hW}{\hat{W}}
\greeknewcommand{\psih}{{\hat{\psi}}}
\greeknewcommand{\psibh}{\hs{1mm}\hat{\hs{-1mm}\psib}{}}
\greeknewcommand{\omh}{\hat{\omega}}
\greeknewcommand{\hK}{\widehat{K}}
\greeknewcommand{\omt}{\tilde{\omega}}
\greeknewcommand{\Omt}{\tilde{\Omega}}
\greeknewcommand{\hh}{\tilde{h}}
\greeknewcommand{\what}{\widehat}
\greeknewcommand{\wti}{\widetilde}
\greeknewcommand{\wGm}{{\widehat{\Gamma}}}

\greeknewcommand\undal{{\underline{\alpha}}}
\greeknewcommand\undel{{\underline{\delta}}}
\greeknewcommand\undbt{{\underline{\beta}}}
\greeknewcommand\undgm{{\underline{\gamma}}}
\greeknewcommand\unddt{{\underline{\delta}}}
\greeknewcommand\undep{{\underline{\epsilon}}}
\greeknewcommand\undsi{{\underline{\sigma}}}
\greeknewcommand\undph{{\underline{\varphi}}}
\greeknewcommand\undSi{{\underline{\Sigma}}}
\greeknewcommand\undOm{{\underline{\Omega}}}

\@ifundefined{goth}{\newfont\goth{eufm10 scaled\magstep1}}{}
\nc{\A}{{\!A}}
\nc{\bA}{{\bar{A}}}
\nc{\bD}{{\bar{D}}}
\nc{\BF}{\xb F}
\nc{\bcd}{\xb\cd}
\nc{\bch}{\overline{\cal H}}
\nc{\bfA}{{\bf A}}
\nc{\bfD}{{\bf D}}
\nc{\bff}{{\bf f}}
\nc{\bfu}{{\bf u}}
\nc{\bfv}{{\bf v}}
\nc{\bfW}{{\bf W}}
\nc{\bh}{{\overline H}}
\nc{\bH}{\xb H}
\nc{\bi}{{\bar{\imath}}}
\nc{\bj}{{\bar{\jmath}}}
\nc{\bk}{{\bar{k}}}
\nc{\bK}{{\bar{K}}}
\nc{\bl}{{\bar{l}}}
\nc{\bp}{{\bar{p}}}
\nc{\bQ}{{\bar{Q}}}
\nc{\bq}{{\bar{q}}}
\nc{\br}{{\bar{r}}}
\nc{\bR}{{\xb R}}
\nc{\bs}{{\bar{s}}}
\nc\bS{\xb S}
\nc\bT{{\hskip.1ex\overline{\hskip-.1ex T}}}%
\nc{\bU}{{\bar U}}
\warnred{bv}{bar V}
\nc\bV{\xb V}
\nc\bW{\xb W}
\nc{\bw}{{\overline{W}}}
\nc{\bcw}{{\overline{\cw}}}
\nc{\bx}{{\overline{X}}}
\nc{\bX}{{\xb{X}}}
\nc{\by}{{\overline{Y}}}
\nc{\bz}{{\bar{z}}}
\nc{\fb}{{\bar f}}
\nc{\ca}{{\cal A}}
\nc{\cb}{{\cal B}}
\warnred{cc}{\cal\space C}\def\cc{{\cal C}}
\nc{\cd}{{\cal D}}
\nc{\ce}{{\cal E}}
\nc{\cf}{{\cal F}}
\nc{\cg}{{\cal G}}
\nc{\ch}{{\cal H}}
\nc{\ci}{{\cal I}}
\nc{\cj}{{\cal J}}
\warnred{ck}{\cal\space K}
\nc{\cl}{{\cal L}}
\nc{\cm}{{\cal M}}
\nc{\cn}{{\cal N}}
\nc{\co}{{\cal O}}
\nc{\cp}{{\cal P}}
\nc{\cq}{{\cal Q}}
\nc{\cR}{{\cal R}}
\nc{\cs}{{\cal S}}
\nc{\ct}{{\cal T}}
\nc{\cu}{{\cal U}}
\nc{\cv}{{\cal V}}
\nc{\cw}{{\cal W}}
\nc{\cx}{{\cal X}}
\nc{\cy}{{\cal Y}}
\nc{\cz}{{\cal Z}}
\nc{\fs}{{\bf f}}
\nc{\rd}{{\rm d}}
\nc{\re}{{\rm e}}
\nc{\rg}{{\rm g}}
\nc{\rA}{{\rm A}}
\nc{\rD}{{\rm D}}
\nc{\shalf}{\sm{1}{2}}
\nc{\undA}{{\underline{A}}}
\nc{\undB}{{\underline{B}}}
\nc{\undC}{{\underline{C}}}
\nc{\prt}{{\partial}}
\nc\LRA{\Longleftrightarrow}
\nc\impl{{\Rightarrow}}
\nc\IMAG{\mathop{\Im\!m}}
\nc\Lie{\mathop{\rm Lie}}
\nc\eln{\mathop{\ell\kern-.1em n}}
\nc\real{\Re\!e\,}
\nc\msh{\mbox{$\frac12$}}
\nc\mynparallel{\not{\!\|\,}}
\nc\sinc{{\rm sinc}}
\nc\tr{\mathop{\rm tr}}
\nc\Tr{\mathop{{\rm Tr}}}
\nc\Div{\partial\cdot}
\nc\dslash{\!{\not\!\partial}}
\nc\dsb{\!{\not\!\bar\partial}}
\nc\FT{{\cal F\!T}}
\nc\mybox{\square}

\nc\gf{{\gamma^5}}
\nc\unity{1\hspace{-0.25em}{\rm l}}

\nc\crc[1]{{\begin{picture}(11,9)(0,1)\put(5.5,4.5){\circle{10}}\put(2,0.8){\elvsf #1}\end{picture}}}%

\nc\bwt[1]{{\overline{\wt{#1}}}}
\nc\lra[1]{\stackrel\leftrightarrow{#1}}
\nc\vect[1]{\stackrel\rightarrow{#1}}
\nc\vecto[1]{\stackrel\longrightarrow{#1}}
\nc\wt{\widetilde}
\nc\xb[1]{{\,\overline{\!#1}}}
\nc{\sm}[2]{\hbox{\footnotesize$\displaystyle\frac{#1}{#2}$}}
\nc\sms[2]{\sm{#1}{#2}\,}
\nc{\thalf}{\tm12}
\nc{\tm}[2]{{\mbox{${#1\over#2}$}\,}}
\nc\bigexp[1]{{\mbox{\large$\rm e^{#1}$}}}
\nc\der[2]{{\partial{#1}\over\partial{#2}}}
\nc{\del}[2]{{{\delta}_{#1}}^{#2}}
\nc\diag[1]{\mathop{\rm diag}\lr(){#1}}
\nc\eqnrf[1]{\eqn{} \rf{#1}}
\nc\hstext[1]{\hs\mbox{#1}\hs}
\nc\idx[1]{\int\!{\rm d}^{#1}x\,}
\nc\inv[1]{{#1}^{-1}}
\nc\ip[1]{\iota_{#1}}
\nc\lr[3]{{\left#1 #3 \right#2}}
\nc\p[1]{{\left(#1\right)}}
\nc\mycase[1]{\left\{\begin{array}{ccl}#1\end{array}\right.}
\nc\mymat[1]{\lr(){\begin{array}#1\end{array}}}
\nc\mydbleqn[3]{\bea\begin{array}{rcl}#1
 \end{array}#2\begin{array}{rcl}#3\end{array}\ena}
\nc\mea{\@ifnextchar[{\marray}{\marray[$\displaystyle\tabskip\z@{\@@@}$&
\global\@eqcnt\@ne\hskip2\arraycolsep\hfil${\@@@}$\hfil&
\global\@eqcnt\tw@\hskip2\arraycolsep$\displaystyle\tabskip\z@{\@@@}$]}}%
\def\marray[#1]#2{\stepcounter{equation}\let\@currentlabel=\theequation
\global\@eqnswtrue\global\@eqcnt\z@\tabskip\@centering
$$\let\@@@=##
\let\\=\@eqncr\halign to \displaywidth\bgroup\@eqnsel\hskip\@centering
\tabskip\z@#1\hfil\tabskip\@centering&\llap{##}\tabskip\z@\cr
#2\@@eqncr\egroup\global\advance\c@equation\m@ne$$\global\@ignoretrue}%
\nc\set[1]{\lr\{\}{#1}}
\nc\sslash{\,/\hspace{-1.5ex}}
\nc\Slash{\,/\hspace{-1.3ex}}
\nc\SLASH[1]{\,/\hspace{-1.#1ex}}
\nc\SU[1]{{\mbox{SU$(#1)$}}}
\nc\un[1]{{\underline{#1}}}
\nc\var[2]{{\delta{#1}\over\delta{#2}}}
\nc\BBB[1]{{\Bbb #1}}
\warnred{I}{implement BBB replacement}
\def\I#1{{\rm I\!#1}}
\nc\BB{\I B}
\nc\CC{\mbox{\rm C\hspace{-0.55em}\sf I~}}
\nc\DD{\I D}
\nc\HH{\I H}
\nc\KK{\I K}
\nc\NN{\I N}
\nc\QQ{{\sf l\hspace{-0.4em}\rm Q}}
\nc\RR{\I R}
\nc\WW{{\sf \backslash\!\!W}}	
\nc\ZZ{{\sf Z\!\!Z}}

\nc\eu[2]{\eps^{#1 #2}\,}
\nc\el[2]{\eps_{#1 #2}\,}
\nc\sig[3]{\sigma^{#1}_{#2#3}}

\nc\dlb{{\crc B}}
\nc\dlf{{\crc F}}

\nc\0{|_{\th=0}}
\nc\ind[1]{_{\rm #1}}
\nc\dk[1]{_{(#1)}}
\nc\uk[1]{^{(#1)}}
\nc\downup[2]{_{#1}^{{\phantom{#1}}#2}}
\nc\du[2]{{}_{#1}{\kern-.1em}^{#2}}
\nc\ud[2]{{^{#1}}{\kern-.1em}_{#2}}%
\nc\dud[3]{{}_{#1}\ud{#2}{#3}}
\nc\udu[3]{{}^{#1}\du{#2}{#3}}
\nc\up[1]{{}^{#1}\!}

\nc\zerobox[2]{{\raisebox{#1}[0pt][0pt]{$\scriptstyle #2$}}}%
\nc\zerozerobox[1]{\zerobox{0pt}{#1}}%

\nc{\lsym}[1]{{\mathop{#1}\limits_{\hbox{\large$\smile$}}}}
\nc{\sym}[1]{{\mathop{#1}\limits_{\scriptstyle\smile}}}
\nc\zsym[1]{ \zerozerobox{  \mathop{#1}\limits_{\zerobox{.5ex}\smile}  } }

\nc\lsy[1]{{%
\setbox\@tempboxa\hbox{$\scriptstyle #1$}
\@tempcnta\wd\@tempboxa
\divide\@tempcnta\unitlength
\@tempcntb\@tempcnta
\divide\@tempcntb by 2\relax
\advance\@tempcnta -4\relax
\begin{picture}(0,0)
\put(\@tempcntb,-1){\oval(\@tempcnta,5)[b]}
\end{picture}
\usebox\@tempboxa
}}

\nc\gp{\up g\Phi}
\nc\gA{ \,\up g\!A}

\nc\vi[1]{\cv\ind{#1}}
\nc\li[1]{\cl\ind{#1}}
\nc\lcin{\li{cin}}
\nc\lmsq{\li{mini~SUSY~QED}}
\nc\lym{\li{YM}}
\nc\lmat{\li{mat}}
\nc\lmass{\li{mass}}
\nc\lpot{\li{pot}}
\nc\lcu{\lcin^{\rm gauge}}
\nc\lcm{\lcin^{\rm mat}}
\nc\lcl{\lcin^{\rm lin}}
\nc\lpm{\lpot^{\rm mat}}
\nc\inth[1]{\int\!{\rm d}^#1\th\,}
\nc\bpp{\lr(){\Phib\Phi}}
\nc\krs{\kappa\dk{r)(s}}
\nc\kp{K(\phi,\phib}
\nc\HHb{\overline\HH}
\nc\Phid{\Phi^\dagger}
\nc\lieco[1]{\pounds_{#1}}	
\nc\UK{{$\rm U_K(1)$}}
\nc\warg{{\{\Phi_k\}}}	
\nc\stw{{\sin\theta_W}} 
\nc\ctw{{\cos\theta_W}} 
\REMARK{
}
\nc\CR{\nonumber\\}
\nc\nn{\nonumber}
\nc\numbertwo{\CR[-1.5ex]\\[-1.5ex]\nn}
\nc\lbl[1]{\label{eq:#1}}
\nc\rf[1]{(\ref{eq:#1})}
\title{The three--form multiplet\\ in $N=2$ superspace}
\author{Maximilian F. Hasler\thanks{allocataire M.\,E.\,S.\,R.; ~
E--mail: \tt hasler@cpt.univ-mrs.fr}
\\
\small
Centre de Physique Th\'eorique\thanks
{Unit\'e Propre de Recherche 7061},
C.N.R.S. Luminy, case 907, F--13288 Marseille}
\def\thesection{\arabic{section}.}
\def\DC{\mbox{\large$\gm$}{\kern-.1ex}}
\begin{document}
\mynewpage
%
%
%
\REMARK{
\def\CPT#1#2{#1}
\pagenumbering{roman}
\begin{titlepage}
\makeatletter
\title{\CPT{}{\vspace*{2cm}}\hbox to 0pt
{The three--form multiplet in $N=2$ superspace}}
\author{Maximilian F. Hasler\thanks{allocataire M.\,E.\,S.\,R.; ~
E--mail: \tt hasler@cpt.univ-mrs.fr
}%
\CPT{\\
\small\em Centre de Physique Th\'eorique\thanks
{Unit\'e Propre de Recherche 7061},
C.N.R.S. Luminy, case 907, F--13288 Marseille}{}%
}
\vfill
}
\date{}
\rightline{CPT--96/P.3349}
\rightline{hep--th/9606076}
\rightline{version March 1997}
{\let\newpage\relax\maketitle}
\begin{abstract}
We present an $N=2$ multiplet including a three--index antisymmetric
tensor gauge potential, and describe it as a solution to the Bianchi
identities for the associated fieldstrength superform, subject to some
covariant constraints, in extended central charge superspace.
We find that this solution is given in terms of an $8+8$ tensor multiplet
subject to an additional constraint.
We give the transformation laws for the multiplet as well as invariant 
superfield and component field lagrangians, and mention possible couplings 
to other multiplets.
We also allude to the relevance of the 3--form
geometry for generic invariant supergravity actions.

\end{abstract}
\REMARK{
\vfill
\vfill
\phantom{Keywords: }gauge theories, supermultiplets.\\[4ex]
\rightline{CPT--96/P.3349}\\[2ex]
{version March 1997}\\[2ex]
\rightline
{hep--th/9606076}\\[3ex]
~\par
anonymous ftp or gopher: cpt.univ-mrs.fr
\vfill
\end{titlepage}
\pagenumbering{arabic}
\mynewpage
}
\section{Introduction}

Although $N=2$ supersymmetric theories are not yet directly 
related to the observable world (at least in what concerns particle physics), 
they started playing an important role in modern theoretical physics in
recent years.  This is in particular due to the fact that they yield
the first exactly solvable models in four dimensional field theories,
in spite of the rich physics they display, including also
non--perturbative phenomena \cite{SW94b}.

In this paper, we discuss an $N=2$ supermultiplet including a
three--index antisymmetric tensor gauge potential. This multiplet
will play a crucial role in various scenarios related to
supersymmetry breaking, in analogy to the case in $N=1$: For
instance a gaugino condensate is associated, from a geometrical
point of view, to the derivative $\rd\cq=\tr\cf^2$ of the Yang--Mills
Chern--Simons form $\cq=\tr(\ca\cf-\tm13\ca^3)$, which is just a
special case of a three--form potential. Similarily, curvature squared
terms, which seem to be involved in another recently discussed
supersymmetry breaking mechanism \cite{HOW95a}, are  associated in the
same way to the gravitational Chern--Simons forms. For $N=1$, the
three--form multiplet has been known for some time \cite{Gat81}, and
recently a rather complete description of its couplings to the $N=1$
supergravity--matter system has been given \cite{BPGG96}.

However, to our knowledge, the corresponding $N=2$ multiplet has not
yet been constructed.  It is the purpose of the present paper to fill
this gap: First, we discuss the field content and the supersymmetry
transformations on the level of component fields, and give an invariant
lagrangian density for the multiplet.
Then, we turn to a geometric description of the multiplet as fieldstrength 
of a three--form gauge potential in extended superspace. We find that this
solution is given in terms of an $8+8$ tensor multiplet \cite{Wess75}
(``linear superfield'') subject to an additional constraint\footnote
{Another ``extra--constrained'' hypermultiplet seems to be known, but it is
apparently impossible to write an action for this multiplet \cite{MF96}.}
similar to the case in $N=1$.
We also give the geometrical interpretation of the previously 
defined supersymmetry transformations.
Finally, we give chiral superfield lagrangians, comment on the dyna\-mics of
this multiplet in various contexts, and mention possible couplings to $N=2$ 
supergravity and to other multiplets.
We also indicate the relevance of the three--form geometry for the
construction of generic invariant actions.

Our spinor notations are those of Wess and Bagger \cite{WeB83}, and
concerning the internal structure group of $N=2$ superspace, we
adopted the conventions of \cite{MM87}: In particular, we raise and lower
internal \SU2--indices from the left by the antisymmetric tensors
$\eps_{BA}$ and $\eps^{BA}$ with $\eps^{12}=\eps_{21}=1$.

\mynewpage
\section{The three--form multiplet}

The set of fields that we will call the three--form multiplet $\Si$ in the
sequel consists of a three--index gauge potential $C_{\ell mn}(x)$,
another two--index antisymmetric tensor $S_{mn}(x)$, an isotriplet of real scalar fields which we write as traceless, hermitean 2$\times$2--matrix
$Z\ud BA(x)$, as well as an isodoublet of Weyl spinors
$\zeta_\al^A(x)$, $\zeb^\da_A(x)$ and a real scalar auxiliary field $H(x)$:
\be
	\Si ~\sim~ \p{ C_{\ell mn},~S_{mn},~Z\ud BA;~
			\zeta_\al^A,~\zeb^\da_A~|~H }
~.\ee

The field tensors
\be
	\wt Z^m = \sms12 \veps^{mnk\ell}\prt_n S_{k\ell}
\com
	\wt\Si = -\prt_m\wt C^m
\ee
(where $C_{k\ell m}=\veps_{k\ell mn}\wt C^n$ etc.) are invariant under the
gauge transformations
\bea
	\dt S_{mn} &=& -\prt_m\beta_n + \prt_n\beta_m	~,
\numbertwo
	\dt \wt C^m &=& -\sms12 \veps^{mnk\ell}\prt_n \DC_{k\ell}
~.\ena

We define the following supersymmetry transformations of
spinorial parameters $\xi^\al_A$ and $\xib_\da^A$ for the multiplet:
\bea
	\dt S_{mn} &=& 2\,\p{\xi_A\si_{mn}\zeta^A + \xib^A\sib_{mn}\zeb_A}
~,\CR
	\dt C_{\ell mn} &=& \veps_{\ell mnk}\,
				\p{ \xi_A\si^k\zeb^A + \xib^A\sib^k\zeta_A}
~,\CR
	\dt Z^{BA} &=& \sum^{BA} \p{ \xi^B\zeta^A + \xib^B\zeb^A }
~,\CR
	\dt\zeta^A_\al &=& \xi_\al^A\p{ i\, \prt^m\wt C_m-H }
\numbertwo&&
	+ \p{ i\,\prt^m Z\ud AB - \dt_B^A\,\wt Z^m } \,(\si_m\xib)_\al^B
~,\CR
	\dt\zeb^\da_A &=& \xib^\da_A\p{ i\, \prt^m\wt C_m + H }
\CR&&
	+ \p{i\,\prt^m Z\ud BA+\dt_A^B\,\wt Z^m} \,(\sib_m\xi)^\da_B
~,\CR
	\dt H &=& i\,\xi_A\dslash\zeb^A - i\,\xib^A\dsb\zeta_A
~.\nn\ena
One can verify that these transformations close on the multiplet:
The commutator $[\dt_2,\dt_1]$ of two supersymmetry transformations of
parameters $\xi_i$ yields a spacetime translation of parameter
\be
	\xi^m = 2i\,\p{ \xi_{2A}\si^m\xi_1^A + \xib_2^A\sib^m\xi_{1A}}
\ee
on all fields of the multiplet, plus gauge transformations with field 
dependent parameters for the potentials, explicitely
\be
	\beta_m = \xi^n S_{nm} + 2i\,\p{
			\xi_{2A}\si_m\xib_1^B + \xib_2^B\sib_m\xi_{1A}} Z\ud AB
\ee
and
\bea
	\DC_{mn} &=& \xi^\ell C_{\ell mn}
	+ 2\,\p{ \xi_{2A}^{}\xi^A_1-\xib_2^A\xib_{1A} } S_{mn}
\CR&&	+ 4\,\p{ \xi_{2A}\si_{mn}\xi_1^B - \xib_2^B\sib_{mn}\xib_{1A} } Z\ud AB
~.\ena

An invariant kinetic action for this multiplet 
is provided by the lagrangian density
\bea
	\li{kin}
&=&
	\tm12 H^2 + \tm12 (\prt_m\wt C^m)^2 + \tm12 \wt Z^m\wt Z_m
\numbertwo
&&
	- \tm14\prt^m Z\ud AB~\prt_m Z\ud BA 
	- \tm i2 \zeta^A\lra\dslash \zeb_A
~.\ena
It should be noted here that $C_{\ell mn}$ does not propagate physical
degrees of freedom. On--shell, the above action describes the same 4+4
physical states as the usual
tensor multiplet~\cite{Wess75}, the 4 physical spin--0 degrees of freedom
being given by the triplet of scalars and the antisymmetric tensor.

Thus, in analogy to the case of the $N=1$ three--form multiplet \cite{Gat81},
the pseudoscalar auxiliary field of the usual matter multiplet is replaced by 
the fieldstrength of the three--form potential.

Actually, the multiplet can be decomposed into one $N=1$ three--form multiplet
and one $N=1$ antisymmetric tensor multiplet \cite{FZW74}:
This is done by writing
\be\lbl{Z-decomp}
	Z\ud BA\sim\pmatrix{ L & ~\bT\cr T & -L}	\com	L=L^\dagger
\ee
and splitting up the $N=2$ multiplet into the two $N=1$ multiplets
\bea
	\Si
~&\leadsto&~~
	(~L,~S_{mn};~\zeta^1_\al,~\zeb_1^\da~)
\numbertwo&&~+~
	(~T,~\bT,~C_{\ell mn};~\zeta^2_\al,~\zeb_2^\da~|~H~)
~.\ena
Then, the supersymmetry transformations of parameter $\xi_\al^1$, $\xib^\da_1$
correspond to the usual $N=1$ transformations of these two multiplets,
while the parameters $\xi_\al^2$, $\xib^\da_2$ mix them in a nontrivial way.
We shall reconsider this issue on the level of superfields after the 
discussion of the superspace Bianchi identities in the following section.

\mynewpage
\section{Superspace geometry and Bianchi Identities}\label{geo-sect}

We find that the previously introduced multiplet can be described as the 
components of the fieldstrength tensor $\Si=\rd C$ of a three--form potential
in extended $N=2$ superspace. We include an additional bosonic coordinate 
$z,~\bz$ which allows for the description of a supersymmetry algebra 
including a central charge \cite{Sohnius:1978}.
Thus, we consider torsion $T^\ca=\rd E^\ca$ 
$(\zerozerobox\ca\sim a,~{^\al_A},~{_\da^A},~z,~\bz)$
with the following nonzero elements:
\be
	T_{\,\gm\,\beta}^{CB~z} = 2i\,\eu CB \eps_{\gm\beta}^{} 
\com
	T_{CB}^{\dg\,\db~\bz} = 2i\,\eps_{CB}^{} \, \eu\dg\db  
~,\ee
and, as usual,
\be
	T_{\,\gm\,B}^{C\,\db~a} = 2i\,\dt^C_B \, (\si^a)\du\gm\db	~.
\ee
By Poincar\'e's lemma $\rd\rd=0$, they define the commutators 
\be
	\p{ \cd_\cc,\big.\cd_\cb } = -T\du{\cc\cb}\ca\cd_\ca	~.
\ee

However, the superfields of the multiplet described here, as well as the
gauge parameters, are taken independent of $z,\bz$,
\be
	\prt_z C_{\cc\cb\ca} = 0
\com
	\prt_\bz C_{\cc\cb\ca} = 0~,
\ee
i.e., the central charge is acting trivially on the multiplet.

Furthermore, we impose the covariant constraints
\be\lbl 0
	\Si_{\un\dt\un{\gm\beta}\ca} = 0
\com
	\Si_{\un\dt\un\gm\,z\bz} = 0
\com
	\p{~{\scriptstyle\un\dt}\sim{_\dt^D},~{^\dd_D}~}
\ee
and
\be\lbl R
	\Si_{z\,c\,\beta\,A}^{~~~B\,\da}
=
	2\,\si\du{c\,\beta}\da\,Z\ud BA
~,~~
	\Si_{\bz\,c\,\beta\,A}^{~~~B\,\da}
=
	-2\,\si\du{c\,\beta}\da\,Z\ud BA	\strut
\ee
on the fieldstrength\footnote
{The latter constraint should be considered as reality condition on $Z\ud BA$.
An antihermitean part of $Z\ud BA$ would just enlarge the field content
of the multiplet by another 8+8 off--shell degrees of freedom.%
}\REMARK
{would yield an additional 8+8 degrees of freedom 
is imposed to restrict the number of independent component fields. 
minimize the number of independent component fields: 
the latter means that $Z\ud BA$ is hermitean: We impose this to reduce 
Relaxing this constraint also yields a consistent solution, with twice the 
number of physical fields, and three auxiliary fields instead of one.}.

Then the Bianchi identities $\rd\Si=0$, or
\be
	E^\ca E^\cb E^\cc E^\cd E^\ce\p { \cd_\ce \Si_{\cd\cc\cb\ca}
	+ 2\, T\du{\ce\cd}\cf \Si_{\cf\cc\cb\ca} } = 0
~,\ee
give all other elements of $\Si_{\cd\cc\cb\ca}$ in terms of $Z\ud BA$ and
its derivatives: First, the identities with indices $z\bz$ imply
\be
	\Si_{z\bz\,\cb\ca} = 0	~.
\ee
(These identities are just the Bianchi identities for a Yang--Mills 
fieldstrength $F_{\cb\ca}$ in ordinary $N=2$ superspace without the
$z$--coordinates \cite{GSW78}, with the Yang--Mills superfields occuring 
in $F_\un{\smash\beta\al}$ set to zero.)

Next, the identities with one index $z$ correspond to
the Bianchi identities $\rd Z=0$ for
\be
	Z_{\cc\cb\ca} \equiv i \Si_{z\,\cc\cb\ca}
\ee
such that $Z=\rd S$ for some 2--form $S$, which is readily identified\footnote
{To be precise, from the reality condition \rf R follows that
$\Si_{z\,\cc\cb\ca}=-\Si_{\bz\,\cc\cb\ca}$, such that
$S'_{\cb\ca} = i\, C_{\bz\,\cb\ca}$ has the same fieldstrength $Z$
and therefore describes the same physical object. (
Roughly speaking,
$S'_{\cb\ca}$ and $S_{\cb\ca}$ differ at most by a gauge transformation.)}
as
\be
	S_{\cb\ca} = -i\, C_{z\,\cb\ca}
~.\ee
These identities imply~\cite{MM87b}
\be\lbl{Z-constr}
	Z\ud AA=0
\com
	\cd_\gm^{(C} Z^{BA)} = 0 = \bcd^\dg_{(C} Z_{BA)}^{}
~,\ee
and give
\be
	Z_{cb\,\al}^{~~A} = 2 \, (\si_{cb}\zeta)_\al^A
\com
	Z_{cb\,A}^{~\,~\da} = 2 \, (\sib_{cb}\zeb)^\da_A
~,\ee
with
\be
	\zeta_\al^A = \tm 13 \cd_\al^B Z\ud AB
\com
	\zeb^\da_A = \tm 13 \bcd^\da_B Z\ud BA
~,\ee
and finally
\be
	Z_{cba} =  \tm 18 \veps_{dcba} \, (\si^d)\ud\al\da \,\p{
		\cd_\al^A \zeb^\da_A - \bcd^\da_A\zeta^A_\al }
~.\ee
As usual, the ``reduced identities'' \rf{Z-constr} already imply that
$\veps^{dcba}\prt_d Z_{cba}=0$.

Then, the other components of $\rd\Si=0$ determine the components of $\Si$
without $z,\bz$ indices to be
\bea
	\Si_{dc\,\beta\,\al}^{~\,~BA} &=& -4\,(\si_{dc})_{\beta\al}Z^{BA}
\com
\numbertwo
	\Si_{dc\,BA}^{~~~\db\,\da} &=& +4\,(\sib_{dc})^{\db\da}Z_{BA} 
~,\ena
and
\bea
	\Si_{dcb\,\al}^{~~\,~A} &=& -\veps_{dcba}\,(\si^a\zeb)_\al^A
\com\numbertwo
	\Si_{dcb\,A}^{~~\,~\da} &=& -\veps_{dcba}\,(\sib^a\zeta)^\da_A
~,\ena
where we chose to adopt the additional constraint 
$\Si_{dc\,\beta\,A}^{\,~~B\,\da} = 0$
which turns out to be conventional, i.e. just corresponds to a shift of the
potential $C_{\ell mn}$ by the trace of this fieldstrength component.
Finally, the vectorial fieldstrength is given by
\bea
	\Si_{dcba} &=& \veps_{dcba} \wt\Si
\com
	\wt\Si = \sms i8 \p{ \cd^\al_A\zeta^A_\al + \cd^A_\da\zeb^\da_A }
~.\ena
On the other hand, one infers of course from the explicit definition of 
$\Si=\rd C$ that
\be
	\wt\Si = -\tm 16 \veps^{dcba} \, \prt_d C_{cba}
~,\ee
therefore the above result should be seen as an additional constraint on the
superfield $Z\ud BA$, requiring the imaginary part of its highest component
to be a total derivative,
\be\lbl{C2}
	\p{ \cd^\al_B\cd_\al^A + \bcd_\da^A\bcd^\da_B } Z\ud BA
=
	\prt_d \p{ 4i\,\veps^{dcba} C_{cba} }
~.\ee
This concludes the analysis of the Bianchi identities; no other
restrictions on $Z\ud BA$ are found. Note that we actually used here 
the same symbols for the superfields than for their lowest components
which are precisely the fields presented in the preceding section.
For convenience, we summarize their definitions here once again, 
denoting the projection on $\th=0$ as usual by a vertical bar:
\be
	Z\ud BA(x) = Z\ud BA \,|
\com
	S_{mn}(x) = -i\, C_{z\,mn} \,|
~,\ee
\be
	\zeta_\al^A(x) = \sms 13 \cd_\al^B Z\ud AB \,|
\com
	\zeb^\da_A(x) = \sms 13 \bcd^\da_B Z\ud BA \,|	
~,\ee
\bea
	\wt Z^m(x) &=& \sms 12 \veps^{mnk\ell} \,\prt_n S_{k\ell}(x)
\CR	&=& \sms 1{24}  (\si^m)\ud\al\da \,\p{
		\bcd^\da_B\cd^A_\al-\cd_\al^A \bcd^\da_B  } Z\ud BA \,|
~,\ena
\bea
\lbl s
	\wt\Si(x) &=& -\sms 16 \veps^{k\ell mn} \, \prt_k C_{\ell mn}(x)
\CR&=&
	\sms i{24} \p{ \cd^\al_B\cd_\al^A + \bcd_\da^A\bcd^\da_B } Z\ud BA \,|
~,\ena
and there remains just one real scalar auxiliary field, defined as
\be
	H(x)
=
	\sms1{24} \p{ \bcd^A_\da\bcd^\da_B-\cd^\al_B\cd^A_\al } Z\ud BA\,|
~.\ee

~

Observe that the additional constraint \rf{C2} is analogous to the case in 
$N=1$: There, the three--form multiplet also corresponds to (anti--)chiral
superfields
\be\lbl{T-constr}
	\bD^\da T = 0	\com	D_\al \bT = 0~,
\ee
satisfying the additional constraint
\be
	D^\al D_\al T - \bD_\da\bD^\da \bT \propto \veps^{dcba} \prt_d C_{cba}
~.\ee
Considering the 
decomposition \rf{Z-decomp} of $Z\ud BA$, we find the
chirality constraints \rf{T-constr} on $T$ from the constraints \rf{Z-constr},
$\cd_\al^{(C}Z^{BA)}=0$, by setting all indices to one, while the linearity
constraints on $L$ and the additional constraint on $T$, $\bT$ 
are most easily recovered from the relations
\bea
	\cd_\beta^B \zeta^A_\al 
&=&
	-\eps^{BA}\,\eps_{\beta\al}^{}\,(H+i\,\wt\Si)
\com
\numbertwo
	\bcd^\db_B {\zeb\vphantom\bcd}^\da_A
&=&
	+\eps_{BA}^{}\,\eps^{\db\da} \,(H-i\,\wt\Si)
~,\ena
again by setting the internal index $\scriptstyle B$ of the spinorial derivative
to one.

~

To conclude this section, we now turn back to
the previously introduced supersymmetry transformations. We define them to
be superspace diffeomorphisms of parameter $\xi^\ca$
such that the vielbeins $E^\ca$ remain invariant, i.e.,
\bea
	\cl_\xi E^\ca \equiv (\ip\xi+\rd)^2 E^\ca 
= 0
&&\numbertwo	
\iff
	\cd_\cb\xi^\ca = - \xi^\cc T\du{\cc\cb}\ca	\lbl{covar-diffeo}
~.\ena
This restricts the parameters to be $x^m$ and $z,\bz$--independent, and
the $\th$ and $\bt$--components of $\xi^a$ and $\xi^{\un z}$ to be given
in terms of $\xi^\al_A$ and $\xib_\da^A$.
We combine these diffeomorphisms with a field dependent gauge transformation
\be
	\dt_\gm C = -\rd \DC
\hstext{with}
	\DC = \ip\xi C
\ee
in order to obtain the very simple transformation laws
\be
	\dt C = \ip\xi\Si
\com
	\dt\Si_{\cd\cc\cb\ca} = \ip\xi\rd\,\Si_{\cd\cc\cb\ca}
~.\ee
The relevant components of these formulae, or rather their values for
$\th=0$, were given in the previous section. (We take the lowest
component of $\xi^m$ and $\xi^\un z$ to vanish for the supersymmetry transformations; this is of course consistent with the condition \rf{covar-diffeo}.)

One can show that the commutator $[\dt_2,\dt_1]$ of two such transformations 
yields again a transformation of the same type, but of parameter
\be
	\xi_{}^\ca
=
	-\ip{\xi_2} \ip{\xi_1} T^\ca
\equiv
	-\xi_2^\cb\,\xi_1^\cc\, T\du{\cc\cb}\ca
\ee
plus a gauge transformation of parameter
\be
	\DC_{} = -\ip{\xi_2} \ip{\xi_1} \Si
\com
	(\DC_{})_{\cb\ca} = -\xi_2^\cc\,\xi_1^\cd\, \Si_{\cd\cc\cb\ca}
~.\ee
Disentangeling furtherly, we can describe this combined
transformation also as a plain diffeomorphism of parameter $\xi$ plus
a gauge transformation of parameter $\DC_{} + \ip{\xi_{}} C$, which
then yields the formulae given in the first section.

~

We also wish to comment on 
the piece in these transformations that stems from the
$z,\bz$--component of the parameter $\xi$: It actually has the form of
something one could view as a kind of ``relic'' central charge transformation,
acting nontrivially only on the three--form potential, which becomes shifted 
by the antisymmetric tensor's fieldstrength,
\be
	\dt_z C_{cba} = \Delta\cdot Z_{cba}
\com
	\Delta = -i\,\p{ \xi^z - \xib^\bz }
~.\ee
However, according to the definition of $Z=\rd S$, this also is nothing else
than a gauge transformation, namely of parameter $\DC =\Delta\cdot S$.

\mynewpage
\section{Superfield lagrangians}

The previously given kinetic action for the three--form multiplet can
actually be obtained as the $\th^4$ component of a chiral 
superfield lagrangian as follows: First, we consider an explicit solution 
to the constraints on $\Si$, namely $C_{z\cb\ca} =i\, S_{\cb\ca}$,
$ C_{\bz\cb\ca} = -i\, S_{\cb\ca}$, with
\bex
	S_{\beta\,\al}^{BA} &=& 4\,\eu BA \eps_{\beta\al}^{} \bS
\com
	S_{BA}^{\db\,\da} = - 4\,\eps_{BA}^{} \, \eu\db\da S
\com
	S_{\beta\,A}^{B\,\da} &=& 0	~.
\ee
Here, $S$ and $\bS$ are (anti--)chiral superfields, in terms of which the
fieldstrength multiplet is then given by
\be
	Z\ud BA = \sms14 \p{ 
		\cd^\vp_A \cd^B_\vp S - \bcd_\dv^B\bcd^\dv_A\bS}
~.\ee
(Note that $S$ and $\bS$ must derive from the same real prepotential, such that
the imaginary part of their highest component is indeed a total derivative
$\sim \prt_m\wt C^m$.)
Then, the previously given kinetic action can be written as
\be
	\li{kin} = \real \inth4 S Z
\com
	Z = -\sms{1}{8} \bcd^A_\dv\bcd^\dv_B Z\ud BA
~.\ee
(Here the chiral volume element is normalized such that $\inth4~\th^4 = 1$.)
Due to the constraints \rf{Z-constr} on $Z\ud BA$, this superfield $Z$ is 
actually a vector superfield, and the above lagrangian is indeed invariant 
under the gauge transformations of $S$~\cite{MM87b}.

We also want to mention that, given this explicit solution, one can add a mass term
\be
	\li{mass} = m^2 \, \inth4 S^2 + h.c.
\ee
which breaks the gauge invariance. (For the usual tensor multiplet, this 
lagrangian has first been considered in \cite{Wess75}).
This yields a theory for a massive multiplet with twice the number of physical
fields:  $\li{mass}$ includes an explicit mass term for a Dirac spinor
made of $\zeta_\al^A$ and $m\cd_\al^A S|$, and for $Z\ud BA$ and the
antisymmetric tensor $S_{mn}$. Moreover, an additional triplet of
bosons, $X\ud BA\sim m(\cd^\vp_A\cd_\vp^B S+\bcd_\dv^B\bcd^\dv_A\bS)$, 
appears as auxiliary fields. Finally, the diagonalization of the fields $H$ and
$\wt \Si$ yields a mass term for $mS$. However, this is a slightly
delicate point:  In complete analogy with the case in $N=1$,
one cannot ``eliminate'' $\wt\Si=-\prt^mC_m$ directly, as it is not
really an auxiliary field. Moreover, its equation
of motion only requires $\wt\Si$ to be a constant, but not to vanish.
As one can see from the supersymmetry transformations, a nonvanishing
constant would give rise to an inhomogeneous transformation law of the
spinor field, which seems to indicate broken supersymmetry.
For a more detailed discussion of these issues, see for example \cite{MFH96}.

\def\uz{\un z}

We wish to mention that the above lagrangians, and even a much larger class
of N=2 invariant actions, find a nice geometric interpretation within the
3--form geometry. Actually, for an arbitrary superfield $L\ud BA$ satisfying
\be
	D_\al^{(A}L^{BC)}_{} = 0
\com
	(\prt_z \mp \prt_\bz) L\ud BA = 0	~,
\ee
we find that
\be\lbl L
%
%
	\cl = \sm{-1}{24}\p{D^\al_A D_\al^B \mp \bD_\da^B\bD^\da_A} L\ud AB
\ee
leads to an invariant action; more precisely
\bea\lbl d
	\dt \cl = \sms i3 \prt_m  	\p{
	\xib^B\sib^m D_A\mp\xi_A\si^m\bD^B} (L\ud AB-\tm12 \dt^A_B\, L\ud FF)
~.\ena
\REMARK{	
\bea\lbl d
	\dt \cl &\equiv& \xi^\un\al D_\un\al \cl = -\tm23 i\, \prt_m X^m|
~,\CR
X^m &=& 
	\p{
	\xib^B\sib^m D_A\mp\xi_A\si^m\bD^B} (L\ud AB-\tm12 \dt^A_B\, L\ud FF)
~.\ena
}%
The formula \rf L applies not only to the case of $z$--independent chiral 
lagrangians ($L\ud AB = D^AD_B X\mp \bD^A\bD_B\bX$ with $\bD^\da_A X=0$),
but can be used to obtain the lagrangians
for the Fayet--Sohnius hypermultiplet \cite{Fay76c,Sohnius:1978}
with \mbox{$L\ud AB=\phib^A(m-\frac i2\lra\prt)\phi_B$}
and for the vector--tensor multiplet \cite{HOW97} 
with $L\ud AB=W^AW_B-\bW^A\bW_B$.


The invariance of the lagrangian \rf L finds a natural explication in the 
3--form geo\-metry: In fact, consider a 4--form 
$L = \tm 1{4!} E^\ca E^\cb E^\cc E^\cd L_{\cd\cc\cb\ca}$ which
verifies $\rd L=0$. Then, a supergravity transformation gives
\be
	\dt L \equiv (\ip\xi + d)^2 L = \rd\ip\xi L	~,
\ee
i.e., $L$ transforms into a total derivative. \nc\te{{\mbox{\normalsize$e$}}}
Projecting on the spacetime components, one finds thus that the lagrangian
\be\lbl l
	\cl = e \,\up*L \equiv \sm \te{24} \veps^{k\ell mn} L_{k\ell mn}(x)
\ee
(with $e=\det e\du ma$) provides an invariant action, since it transforms into
\be
	\dt\cl=\prt_m \p{\sms \te6\veps^{k\ell mn}\xi^\un\al L_{\un\al k\ell n}}
~.\ee
As we showed, the solution of $\rd L=0$ subject to constraints \rf 0 
gives all components of $L$ in terms of 
$L_{\uz\,c\,\beta A}^{~~~B\da} = 2i\,\si_c\du\beta\da L_\uz\ud BA$ 
and its derivatives. (Eq.~\rf l reduces in the flat case to eqs.~\rf L 
resp.~\rf s for appropriate choices of $L_\uz\ud BA$).
This provides therefore a generic method to construct invariant
lagrangians \rf l from arbitrary ``linear superfields'' $L_\uz\ud BA$.
(An similar remark in the case of $N=1$ supergravity lagrangians has been 
made in \cite{GG91}.)
In view of the large amount of necessary definitions and technicalities,
we leave the details of the calculations in $N=2$ supergravity and a more 
elaborate discussion of this issue to a separate publi\-cation.

\mynewpage
\section{Couplings to other multiplets}

As the 3--form gauge multiplet is a special case of a tensor multiplet,
it allows for self--couplings and couplings to other multiplets in the 
same way than they do. 
Notably one could consider the analogue of the action for the improved tensor 
multiplet \cite{WR82}, which can be coupled to a general $N=2$ supergravity
background \cite{WPP83}. For this multiplet, one also can add a mass term 
involving only the invariant fieldstrength multiplet.
The massive three--form multiplet should actually be of interest mainly in the 
supergravity--coupled case.

On the other hand, another subject of special interest related to the
three--form multiplet are of course couplings to other gauge multiplets that 
can be formulated on geometric grounds. In analogy to $N=1$ 
\cite{Gri94,MFH96} we consider, for example, a modified field tensor
\be
	\Si = \rd C + \kappa\,H\wedge A	~,
\ee
where $H=\rd B$ is the fieldstrength of a two--form potential and $A$ is a 
U(1) gauge potential. This gives rise to the modified Bianchi Identity
\be
	\rd\Si = \kappa\,H\wedge F	\com	F=\rd A	~,
\ee
which yields in particular a modified fieldstrength $\wt\Si$, including the
spinorial superpartners of $A_m$ and $B_{mn}$.
This leads to dynamics similar to the case in $N=1$, with notably quartic 
potential terms for the spinor fields, which might again be of interest in 
the scenarios alluded to in the introduction.
These topics
are still to be worked out in detail and will be discussed elsewhere.

\section*{Acknowledgements}

I should like to thank R.~Grimm for introducing me to supersymmetry and for 
useful suggestions during the course of this work.

\mynewpage
\appendix

\mynewpage

\def\bibname{References}
\addcontentsline{toc}{section}{\bibname}


\end{document}